# Bioinspired molecular qubits and nanoparticle ensembles that could be initialized, manipulated and readout under mild conditions

Mingfeng Wang[*a], Yipeng Zhang[a], Wei Zhang[a]

[a] School of Science and Engineering, The Chinese University of Hong Kong, Shenzhen, Guangdong, 518172, P. R. China

[*] Corresponding author. Email: wangmingfeng@cuhk.edu

**Abstract**

Quantum computation and quantum information processing are emerging technologies that have potential to overcome the physical limitation of traditional computation systems. Present quantum systems based on photons, atoms and molecules, however, all face challenges such as short coherence time, requirement of ultralow temperature and/or high vacuum, and lack of scalability. We report a new type of molecular qubits and nanoparticle ensembles based on thermally controllable transformation between *J*-aggregation and monomeric states of molecular chromophores, using pyrrolopyrrole cyanine tethered with polymeric chains such as polycaprolactones as an example. Such supramolecular quantum systems, resembling some feature of light harvesting complexes in photosynthesis, provide new opportunities for manipulating quantum information under mild conditions, which do not require complicated ultra-cooling and/or high vacuum often involved in present superconducting qubits or Rydberg atoms for quantum computation and information processing.

## 1. Introduction

Quantum computation that implements properties of quantum materials and power of information technology has emerged as a potentially revolutionary technology for next generation of computation and information processing[1-3]. Distinct from the "0" and "1" bits in conventional computer systems, quantum computation relies on qubits that represent a linear combination, also called superposition, of "0" and "1" states of a physical system. Present physical systems that have been explored as qubits for quantum computation mainly include photons, superconducting circuits, trapped atoms/ions, quantum dots, nuclear spins and electron spins[4].

Despite the enormous progress on the development of physical systems towards quantum computation, very few of them could meet DiVincenzo's all five criteria for construction of a quantum com-puter: (1) a scalable physical system with well characterized qubits; (2) the ability to initialize the state of the qubits to a simple fiducial state; (3) long relevant decoherence times, much longer than the gate operation time; (4) a "universal" set of quantum gates; (5) capability for qubit-specific measurement[5]. For instance, trapped ions as qubits have shown high fidelities due to their long coherence times (longer than one second) for quantum operation, but require ultralow temperature and ultrahigh vacuum to minimize the environmental interference[6-9]. There are also challenges in scaling up trapped ions into qubit arrays with well-defined connections. Quantum computation based on superconducting circuits has also experienced significant progress from basic research to practical demonstration. For instance, D-Wave has demonstrated a 16-bit superconducting quantum computer, and fabricated computation circuits involving 28 superconducting quantum bits. More recently, quantum walks on a two-dimensional 62-qubit superconducting processor were reported by Pan and coworkers[10]. Nevertheless, superconducting quantum computation still faces challenges such as noise interference, relatively weak coherence and requirement of ultracold cooling system. In addition, gate-controlled quantum dots have been proposed for quantum computation by controlling the electron spins confined within each quantum dot[11]. But it remains challenging to localize single quantum dot precisely between pre-defined gates and scale up further toward large-scale integration. Quantum electrodynamics (QED) involving optical cavity and neutral gas-state atoms also shows potential for quantum computation[12]. Technical challenges existing in QED include improvement of the quality of the optical cavity, integration of large arrays of optical cavity onto circuits, and exploration of new atomic or molecular systems with strong optical coupling and long coherence time.

Our research interest focuses on exploration of new molecular and supramolecular quantum systems inspired by nature, such as light-harvesting complexes in photosynthesis that enables quantum information processing under mild conditions. There has been increasing evidence to support the hypothesis that quantum effect plays a certain role in biological systems[13, 14]. For instance, using ultrafast spectroscopy, Fleming and coworkers[15] observed that quantum superposition and coherence dynamics are related to the ultrafast dynamics of excited state in process of energy transfer and charge separation. Moreover, persistence of quantum coherence at physiological (room) temperature has been observed by Scholes and coworkers in a light-harvesting protein from marine algae[16].

Herein we report a new quantum system composed of multiple molecular chromophores as qubits that are confined in colloidal nanoparticles. Each of the molecular chromophores consists of a π-conjugated unit decorated with flexible hydrocarbon chains. Such molecular design enables not only the well dispersion of the chromophores in matrices such as polymers, but also the reversible and thermally controllable coupling among chromophores. The coupling and the superposition among chromophores, resembling the *J*-aggregates of chlorophylls supported by protein scaffolds in light harvesting complexes of photosynthetic systems, depend on the nature of the matrices and the external environment such as temperature. On the one hand, when the temperature is above the threshold of the phase transition, the chromophores exist as noncoupled monomeric state (denoted as "0" quantum state). On the other hand, when the temperature is below a certain threshold, the intermolecular coupling among chromophores is enhanced and the chromophores all exist as coupled state (denoted as "1" state). In the temperature region between these two thresholds, the chromophores exist as a linear combination of "0" and "1" states, also called as superposition states. The "0" and "1" states correspond to distinct energy potential which are measurable via optical spectroscopy. Such supramolecular qubits open up new opportunities for scalable quantum computation and quantum information processing under mild conditions that do not require complicated systems for ultracold cooling and/or ultrahigh vacuum.

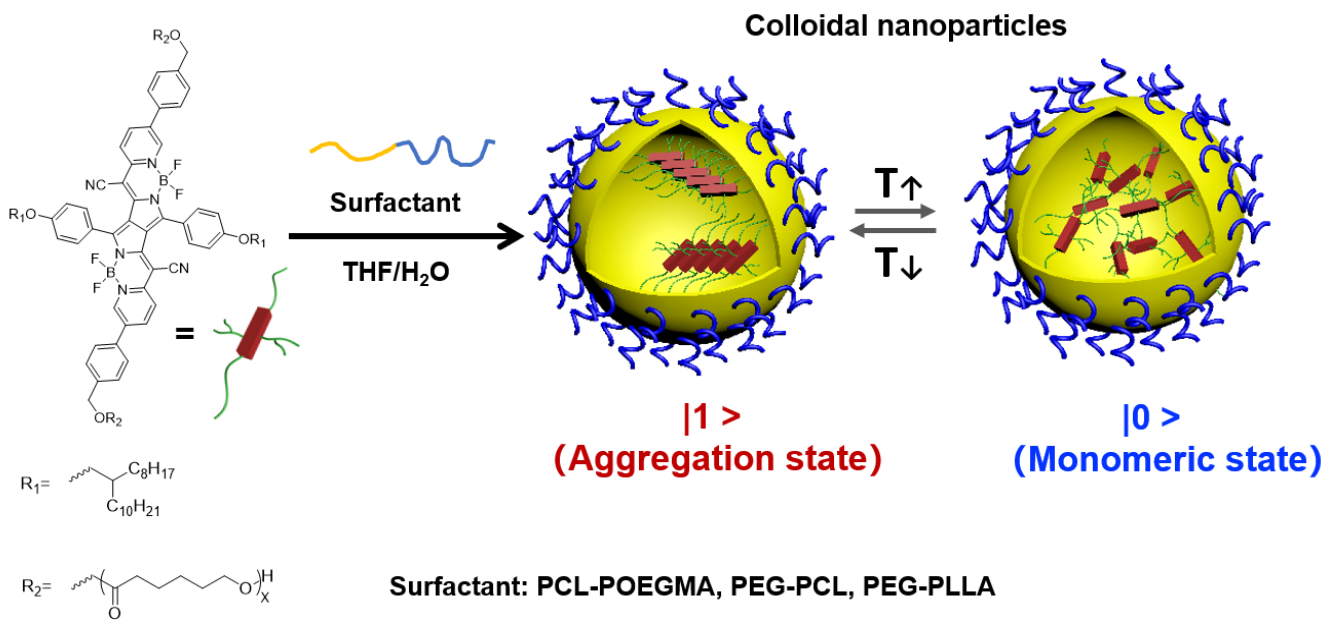

Scheme 1. **Schematic presentation of chemical structure of pyrrolopyrrole cyanine (PPCy) chromophore as molecular qubits,** the preparation of colloidal nanoparticles in the presence of an amphiphilic block copolymer as surfactant, and temperature-induced reversible transformation between J-aggregation state ("1") at a low temperature and monomeric state ("0") at a high temperature. (x represents the number-average degree of polymerization)

## 2. Results and discussion

A representative molecular chromophore reported here consists of pyrrolopyrrole cyanine (PPCy) as a rigid planar core ("plate"), from which flexible coils of benzoxy alkyl and oligo- or polyesters are tethered at different positions. (Scheme 1) The chain length of polyesters is tunable through controllable ring-opening polymerization. Such coil-plate-coil molecules are denoted as PPCy-n, where n represents the average number of repeating units in polyesters. PPCy-n are intrinsically hydrophobic and therefore insoluble in polar solvents such as alcohols and water. As a consequence, PPCy-n chromophores, when n ≤ 25, automatically form colloidal J-aggregates when co-precipitating in water in the presence of a surfactant such as an amphiphilic diblock copolymer poly(ethylene glycol)$_{113}$-b-poly(caprolactone)$_{52}$ (PEG$_{113}$-PCL$_{52}$), poly(caprolactone)$_{37}$-b-poly(oligoethylene glycol

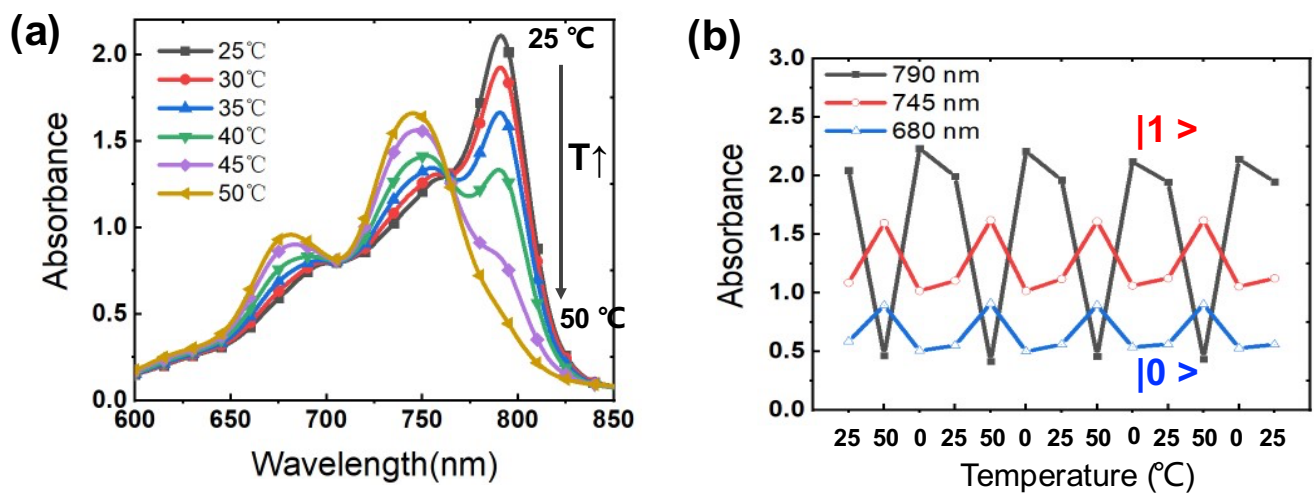

Figure 1. **Temperature-responsive optical properties of colloidal NPs of PPCy-15 stabilized with PCL$_{37}$-POEGMA$_{90}$.** (**a**) Absorption spectra of colloidal NPs of PPCy-15 at different temperatures; (**b**) Absorbance changes of colloidal NPs of PPCy-15 at 790, 745 and 680 nm during 4 heating-cooling cycles.

methacrylate)$_{90}$ (PCL$_{37}$-POEGMA$_{90}$), or poly(ethylene glycol)$_{66}$-*b*-poly(lactide)$_{22}$ (PEG$_{66}$-PLA$_{22}$). The population of *J*-aggregates versus that of PPCy-n monomers, both of which are confined within the colloidal nanoparticles (NPs) with an average diameter of 30-60 nm, increases with the decrease of the value of n. In the present article, we take PPCy-15 as an example to illustrate the concept of supramolecular qubits.

Figure 1 shows the UV-VIS-NIR absorption spectra of colloidal NPs of PPCy-15 at different temperatures in water. One can observe that, at 25 °C, a sharp absorption peak appears at 788 nm, which corresponds to the *J*-aggregation state of PPCy units. At the same time, two small shoulder peaks appear at 679 and 744 nm, respectively, which correspond to the monomeric (non-aggregating) state of PPCy. (Figure 1a) These results suggest that *J*-aggregation remains the dominating state of PPCy inside the colloidal NPs of PPCy-15 at 25 °C. The population of the monomeric state of PPCy could be further minimized by either reducing the chain length of the PCL tethered from PPCy or decreasing the temperature or both. When the colloidal dispersion of PPCy-15 NPs was heated to 50 °C, one can see that the light absorption peak at 788 nm disappeared completely, while the absorbances at 679 and 744 nm were both enhanced significantly. (Figure 1a) These results indicate the complete dissociation of *J*-aggregation into monomeric state of PPCy at 50 °C.

Importantly, the relative population of the *J*-aggregation vs. that of the monomeric state of PPCy inside the colloidal NPs could be precisely tuned by temperature. Figure 1a shows that each temperature in the range of 25-45 °C corresponds to a distinct linear combination of the two states, represented by different absorbances at 679, 744 and 788 nm, respectively. In order to understand the thermal equilibrium of the two states of PPCy inside the colloidal NPs, we further monitored the evolution of the absorption spectra during the heating process from 25 to 35 and 35 to 40 °C, respectively. One can see that, the absorbance at 788 nm underwent a fast decrease in the initial 10 min of heating from 25 to 35 °C, followed by a slow decrease until reach of a plateau. (Figure 2a, 2d) With the further increase of the temperature from 35 to 40 °C, a similar trend of kinetics can be observed from both the decrease of the absorbance at 788 n m and the increase of the absorbance at 752 nm (Figure 2b-c). These results demonstrate that the linear combination of the *J*-aggregation ("1" quantum state) and the monomeric state ("0" quantum state) is a function of temperature, which enables precise control of these two quantum states in a facile way.

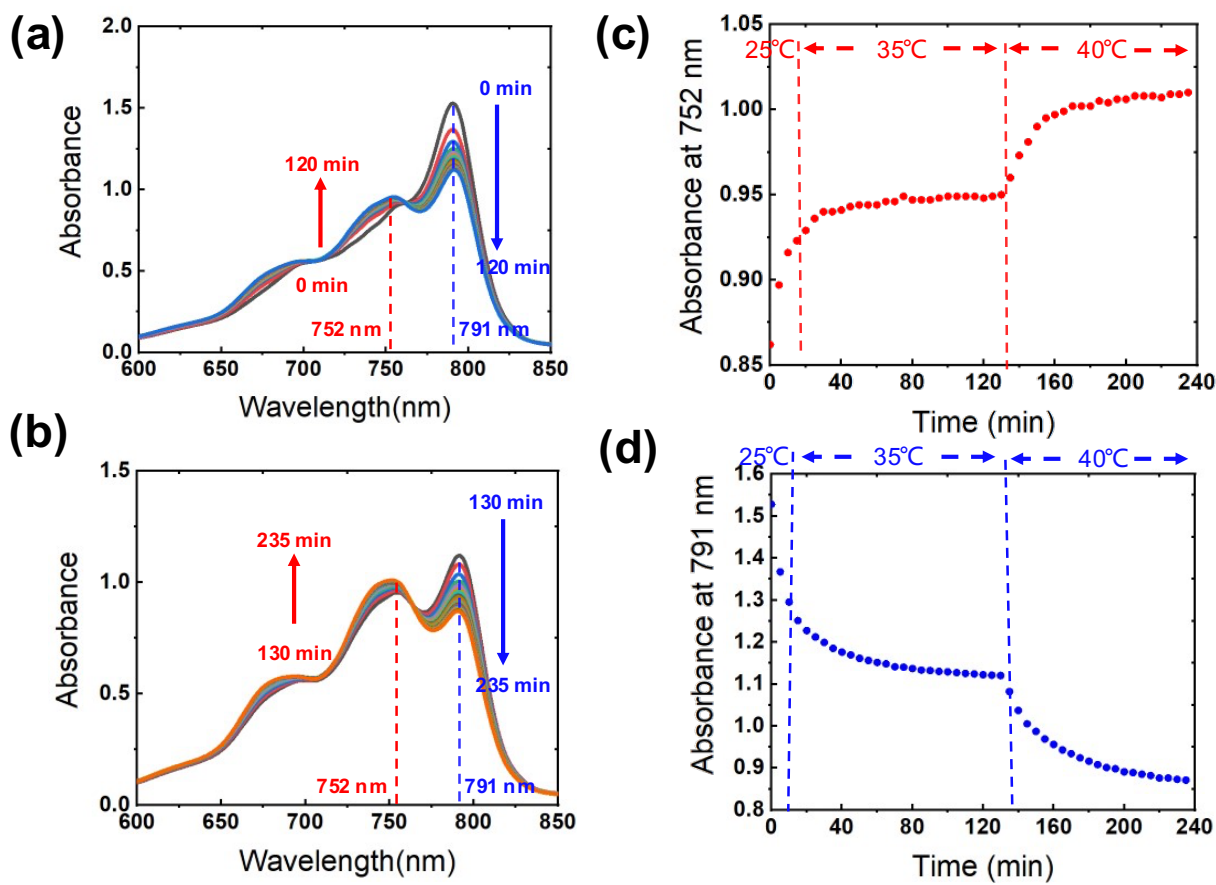

Figure 2. **Kinetics of colloidal PPCy-15 NPs heated at a certain temperature over different periods.** Absorption spectra of colloidal NPs of PPCy-15 in the presence of PCL$_{37}$-POEGMA$_{90}$ heated from 25 to 35 °C (**a**) and 35 to 40 °C (**b**); Absorption changes of colloidal NPs of PPCy-15 at 752 nm (**c**) and 791 nm (**d**).

We further characterized the reversibility of the temperature-induced transformation between the *J*-aggregation and the monomeric state of PPCy-15 NPs. The aqueous dispersion of PPCy-15 NPs was heated from 25 to 50 °C (10 °C/h), followed by a cooling (10 °C/h) process to 0 °C, and then warmed up to 25 °C. The whole heating/cooling process was repeated by four cycles. The results of UV-Vis-NIR absorption spectra show that the transformation between the *J*-aggregation at 25 °C and the monomeric state at 50 °C is fully reversible. (Figure 1b) In addition, the population of *J*-aggregation formed at 0 °C is slightly higher than that at 25 °C.

The temperature-dependent transformation between *J*-aggregation and the monomeric state of PPCy-15 NPs could also be characterized using fluorescence spectroscopy. One can see that with the increase of the temperature from 25 to 40 °C, the intensity of the fluorescence (FL) peak at 800 nm decreased gradually. (Figure 3) Meanwhile, there was little change of the maximum FL emission wavelength, but a small tail of FL emission in the region of 720-770 nm appeared at 35 °C and the FL intensity of the shoulder further increased at 40 °C. When the sample was heated to 45 °C, the FL emission peak was obviously broadened, and the FL intensity was further reduced in comparison with that at 40 °C. Interestingly, with the further increase of the temperature from 45 to 50 °C, there is no significant change of the FL intensity, whereas the wavelength of the maximum emission shifted to 770 nm that corresponds to the FL emission of PPCy monomers. These results indicate that the quantum states of PPCy-15 NPs at different temperatures can also be characterized and monitored by fluorescence spectroscopy. Similar results were obtained when the polymeric surfactant was changed to PEG$_{66}$-PLA$_{22}$ (Figure S2).

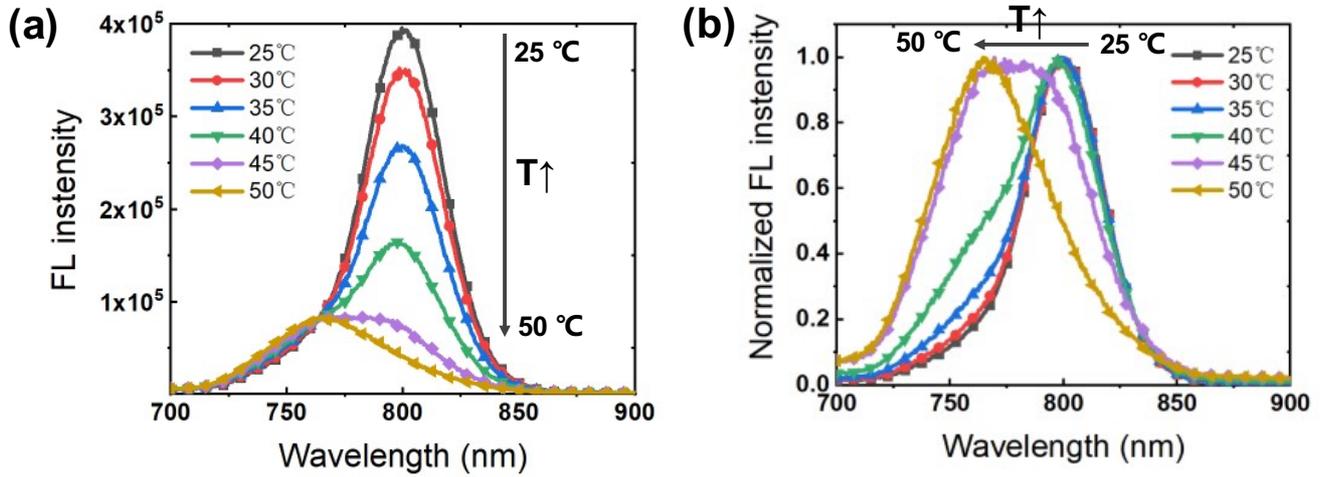

Figure 3. **Steady-state fluorescence emission spectra of colloidal NPs of PPCy-15.** Fluorescence emission spectra (**c**) and normalized fluorescence emission spectra (**d**) of colloidal NPs of PPCy-15 in the presence of PCL$_{37}$-POEGMA$_{90}$ at different temperatures ($\lambda_{ex}$ = 650 nm).

The aqueous dispersion of the PPCy-15 NPs as described above are amenable to be processed into thin films in the presence of polymeric additives such as poly(vinyl alcohol) (PVOH) as matrices. The NPs dispersed in PVOH matrices also show temperature-responsive optical properties. With the increase of temperature from 25 to 55 °C, the absorption peak at 788 nm was attenuated gradually, accompanied with increase of the absorbance at both 679 and 744 nm. (Figure 4) Nevertheless, compared to the colloidal NPs dispersed in aqueous media, the same NPs dispersed in solid PVOH matrices show a higher threshold of temperature, i.e. latency, in the thermally induced transition from *J*-aggregation to monomeric state. These results suggest the possibility of manipulating the quantum state of PPCy in solid films that are scalable and amenable to integrate with other optoelectronic devices.

The present molecular qubits and their ensembles confined in matrices of polymeric nanoparticles represent a new member in the family of many-body quantum physics. Traditional quantum systems in many-body physics, such as ultracold atoms (also called Rydberg atoms) and molecules (largely diatomic systems such as KRb), often requires ultralow temperatures to cool the gas-state atoms or molecules, in order to control the motional degrees of freedom and thus the energy potential of the quantum states[17, 18]. Distinct from the traditional many-body quantum systems, our nanoparticular ensembles of molecular qubits could be initialized to "0" (monomeric state) by heating to a temperature (e.g. 50 °C) above the phase transition temperature. And the superposition or quantum entanglement among molecular qubits, represented by interplay between monomeric state and the *J*-aggregation coupling of chromophores, could be finely controlled by cooling to a certain temperature, for example, in the range of 25-50 °C. Therefore, the present nanoparticles composed of multiple molecular chromophores as qubits show significant advantages over traditional many-body quantum systems, such as no requirement of ultralow temperature and/or high-vacuum, strong intermolecular coupling against decoherence and ease of scalability.

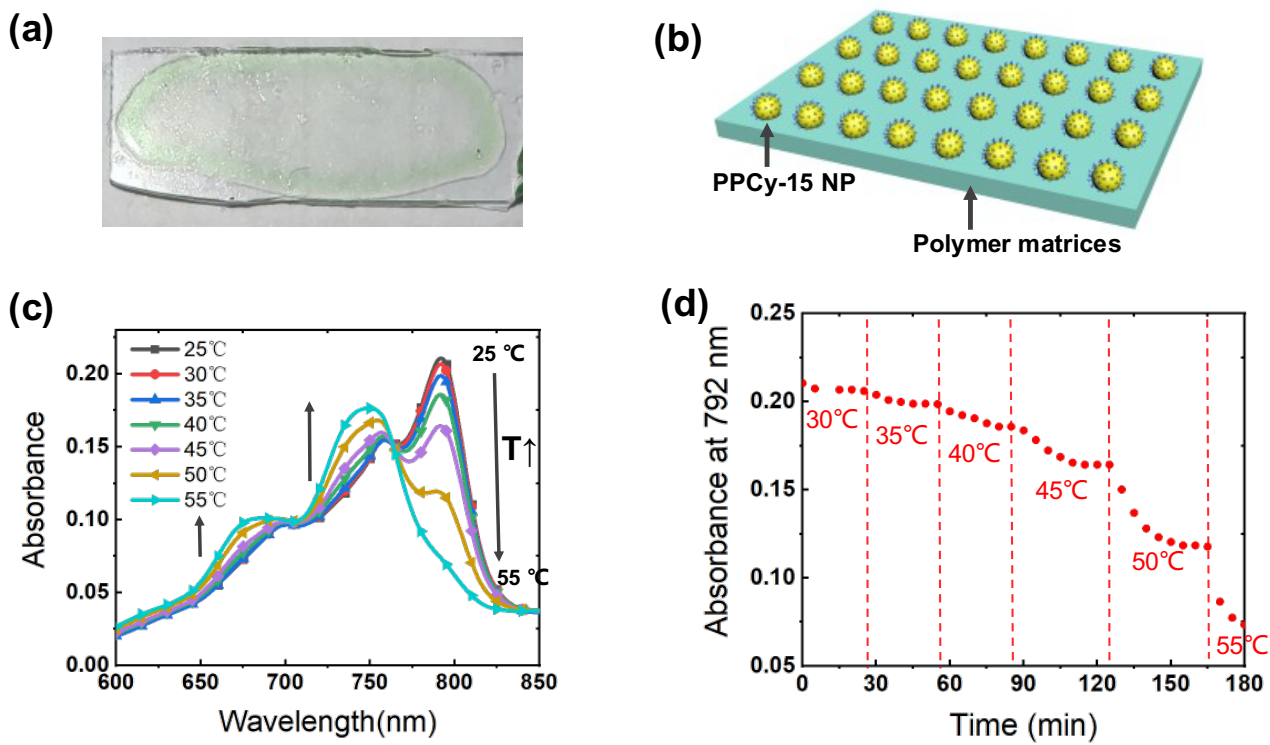

Figure 4. **Temperature-responsive optical properties of PPCy-15 NPs in films.** A representative digital photograph (**a**) and schematic presentation (**b**) of PPCy-15 NPs dispersed in matrices of PVOH films; (**c**) Absorption spectra of PPCy-15 NPs in films at different temperatures; (**d**) Evolution of the absorbance at 792 nm of PPCy-15 NPs in films heated at a certain temperature over different periods.

## 3. Conclusion

We have presented a molecular quantum system composed of polymer-tethered chromophore of a pyrrolopyrrole cyanine derivative. The precipitation of the chromophore in a mixed solvent of tetrahydrafuran and water in the presence of an amphiphilic block copolymer as surfactant, followed by removal of the organic vapor, results in colloidal nanoparticles of chromophore ensembles well dispersed in polymeric matrices. The colloidal nanoparticles undergo reversible phase transformation between *J*-aggregation at a relatively low temperature (0-25 °C) and monomeric state at a relatively high temperature (> 50 °C). The *J*-aggregation coupling between the chromophores confined within colloidal nanoparticles could be precisely controlled by temperature. The quantum states of *J*-aggregation and monomers are clearly distinguishable and measurable using optical spectroscopy. Such concept of molecular qubits and many-body supramolecular quantum systems could be applicable to a broad range of chromophores in which their aggregation states are controllable and distinguishable from the monomeric states.

## Conflicts of interest

There are no conflicts to declare.

## Acknowledgements

We acknowledge the financial support by the University Development Fund – Research Start-up Fund (UDF01001806) from the Chinese University of Hong Kong, Shenzhen. We thank the Advanced Materials Laboratory and the School of Science and Engineering for access to the instrumental platform such as spectrometers and gel-permeation chromatography. We also thank the School of Life and Health Sciences for access to the fluorometer. Y.Z. and W.Z. thank the financial support of Ph.D. Scholarship from the Chinese University of Hong Kong, Shenzhen.

*Mod Phys* **2003,** *75* (2), 457-472.

# Supporting Information

## Materials and Characterizations

### Materials
Chemicals were used as received unless otherwise indicated. Surfactants and PPCy-15 were synthesized in previous work[1-3]. Poly(vinyl alcohol) (89-90% hydrolyzed, average $M_n$ = 30000-70000 g/mol) was purchased from Sigma-Aldrich and used as received.

### Characterization
UV/Vis/NIR spectra of the samples were measured with a Shimadzu UV-1900 spectrophotometer. Fluorescence emission spectra were acquired using a fluorometer (Horiba Fluorolog-3).

### Preparation of nanoparticles in the presence of surfactant as stabilizer.
Colloidal nanoparticles with 0.4 mM as representative example, PPCy-15 (0.2 mL of a 0.01 M THF solution) and amphiphilic co-polymer (PCL$_{37}$-POEGMA$_{90}$, or PEG$_{113}$-PCL$_{52}$, or PEG$_{66}$-PLA$_{22}$, 0.1 mL of a 5 g L$^{-1}$ THF solution) were mixed in 0.2 mL of THF. This mixture was rapidly injected into 5 mL of deionized water under ultrasonication. THF was evaporated by exposure to open air overnight.

### Preparation of nanoparticles films in the presence of poly(vinyl alcohol) (PVOH) as matrices.
PPCy-15 nanoparticles (0.4 mM aqueous solution, 0.5 mL) and poly(vinyl alcohol) ($M_n$ = 30000-70000 g/mol, 0.1 g L$^{-1}$ in water, 0.5 mL) were mixed in a centrifuge tube. This mixture was dropped onto a glass substrate and the film was dried in the open air overnight to form a thin, transparent film.

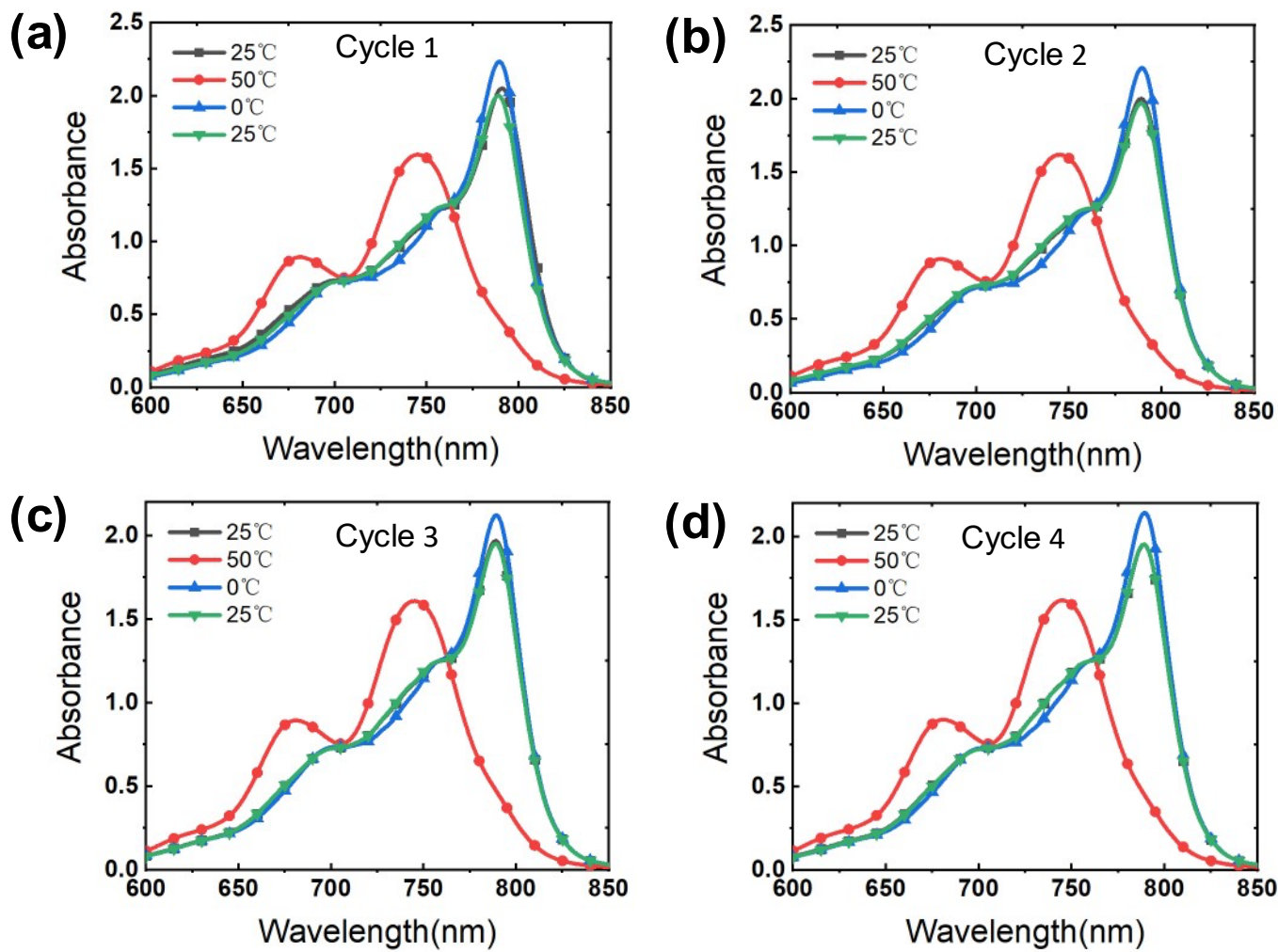

Figure S1. **Absorption spectra of colloidal NPs of PPCy-15** stabilized by PCL$_{37}$-POEGMA$_{90}$ in aqueous media during 4 cycles of heating-cooling.

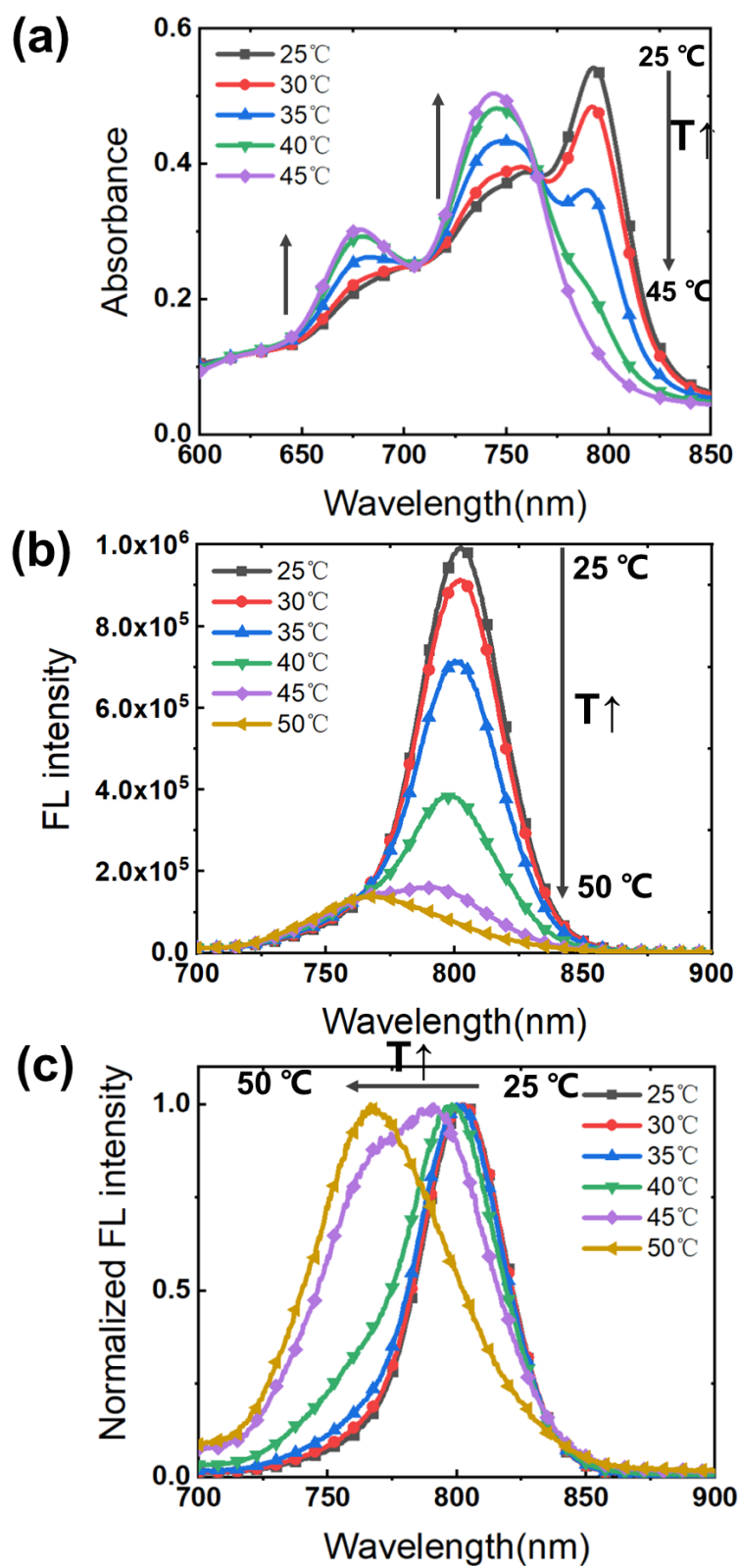

Figure S2. **Temperature-responsive optical properties of colloidal NPs of PPCy-15** stabilized by PEG$_{66}$-PLA$_{22}$ in aqueous media. (**a**) Absorption spectra; (**b**) Fluorescence emission spectra; and (**c**) Normalized fluorescence emission spectra at different temperatures ($\lambda_{ex}$ = 650 nm).

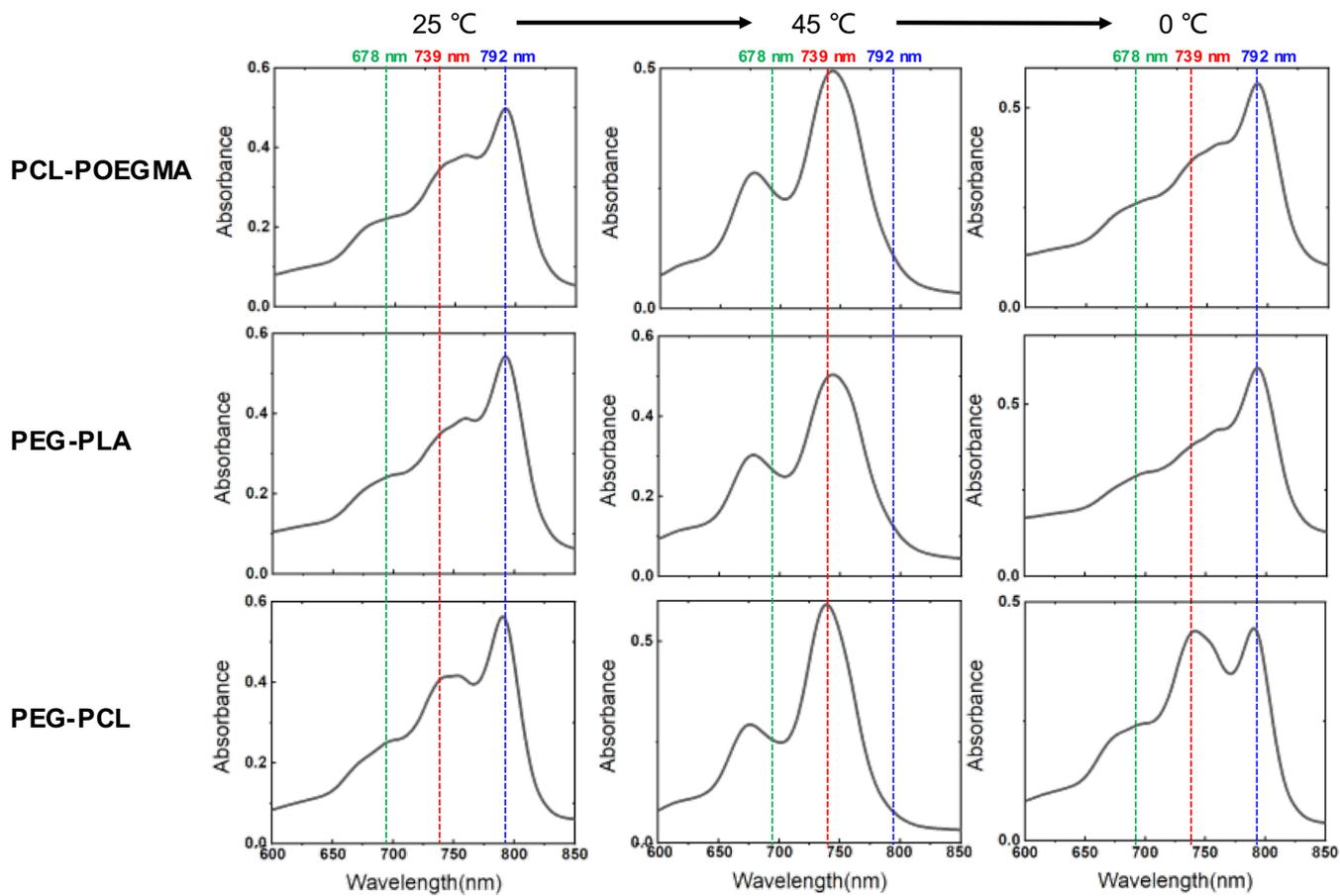

Figure S3. **Absorption spectra of colloidal NPs of PPCy-15** in the presence of different surfactants (PCL$_{37}$-POEGMA$_{90}$, or PEG$_{113}$-PCL$_{52}$, or PEG$_{66}$-PLA$_{22}$) at different temperatures.